# On Boosting the Throughput with Minimal Emitted No. of Molecules for the Diffusion-Based Molecular Communication networks: Prospective and Challenges

Ahmad Yousef[1]

[1]School of Computational Science and Engineering, McMaster University, Hamilton, Ontario, Canada

*Correspondence: mohamas2@mcmaster.ca

***ABSTRACT***- It is assumed that each nano-robot has a 0.1µm3 tankage which is far smaller than the volume of a human red blood cell (i.e. 1.12× 102 $\mu m^3$). Thus, the nano- machines are non-invasive for biomedical applications, such as drug delivery and disease diagnosis. Since the nano- machine has an initial source of 108 molecules in its own tankage and since the throughput of the DMC systems is significantly low; we propose and discuss some ideas to boost the throughput with emphasize of reducing the emitted no. of molecules per message.

## I. Background & Related Works

The use of molecules, instead of electromagnetic or acoustic waves, to encode and transmit the information represents a new communication paradigm that requires novel solutions such as molecular transceivers, channel models or protocols for nanonetworks. Diffusion-based communication refers to the transfer of information using molecules as message carriers whose propagation is based on the law of molecular diffusion [1]. In nanonetworks, we can identify five different components: the transmitter node, the receiver node, the messages, the carrier, and the medium. Each of these components affects the overall communication process, which includes the following steps: The transmitter encodes the message onto molecules; the transmitter inserts the message into the medium by releasing the molecules to the environment or attaching them to molecular carriers; the message propagates from the transmitter to the receiver; the receiver detects the message; and the receiver decodes the molecular message into useful information such as reaction, data storing, actuation commands, etc. Pulse shaping is considered one of the unexplored challenges in molecular communication systems until this moment. In fact, according to Brownian motion, the molecular channel response has a long tail which causes residual noise and Inter Symbol Interference (ISI) in the communication channels [1], [5]. In order to minimize these problems, similar to classical communication systems, it is suggested that a pulse shaping filter be used at the transmitter to minimize the ISI impact. Channel characteristics for the propagation medium must be studied in further detail before an appropriate pulse shaping filter can be selected. In [7], two modulation techniques for the Diffusion-based molecular communication networks are proposed. The first scheme Concentration Shift Keying (CSK) modulates the information via the variation in the concentration of the messenger molecules whereas the second scheme, MoSK, utilizes different types of messenger molecules to represent the information. The names of derivatives of the modulation schemes are according to the number of bits per symbol as BCSK, QCSK, BMoSK, and QMoSK techniques, respectively. In Binary Concentration Molecule Shift Keying (BCSK) scheme, the transmitter emits the molecules in instantaneous fashion. At the reception node, logical "1" is represented by certain threshold no. of molecules; otherwise it's logical "0". CSK technique can be affected adversely from Inter Symbol Interference (ISI) which can be caused by the surplus molecules from previous symbols. Due to the diffusion dynamics, some messenger molecules may arrive after their intended time slot. These molecules cause the receiver to decode the next intended symbol incorrectly. Similar to the CSK technique, the residual molecules from the previous symbols also cause ISI when MoSK technique is used. However, MoSK scheme is less sensitive to ISI effects than the CSK technique. This advantage of the MoSK technique comes at the cost of the requirement for complex molecular mechanisms at both the transmitter and the receiver for messenger synthesis and decoding purposes, respectively. In [4], an optimum receiver is proposed for MoSK technique in case of a linear and time invariant signal propagation model and an additive noise model for the DMC channel. Regarding medium access control (MAC) protocols, it is noticeable that the high propagation delay in DMC systems limits the use of carrier sensing (CSMA) [2]; e.g., at a given time instant, the transmitter may sense a channel free of molecules, but the medium could be very busy at the receiver for various possible scenarios.

location. An alternative MAC protocol proposed for DMC is molecular-division multiple access (MDMA) [3]. This technique uses different molecule types in order to perform several simultaneous transmissions, sharing the same medium but without interfering with each other. Some concerns regarding this technique include the choice of molecule types to be used by different transmitters so that they do not interact among them and receivers are able to recognize each of the received molecules from one another.

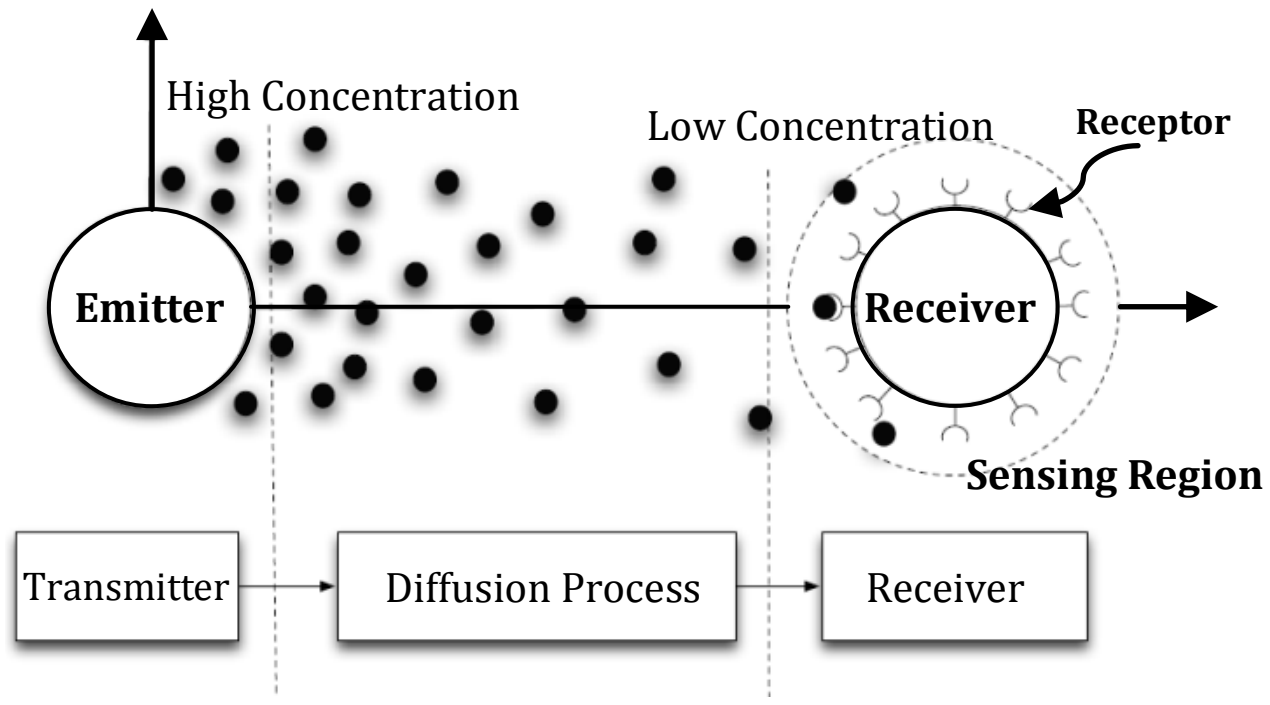

**FIG 1, The Communication Process According to Brownian Motion**

The rest of this paper is organized as follows. In Section II, we propose Disciplined Peak process. Channel capacity maximization is presented in Section III. Moreover, Multi-Hopping Gain is discussed in Section IV. And finally, Section V concludes our work.

## II. Disciplined Peak Process

Disciplined Peak (DP) process controls or reduces the transmitted signal Q based on the previous transmitted signals so that the peaks of the actual received signals will approximately remain constant at the receiver. This process has the ability to suppress the effect of ISI so that the data rate can be boosted without high BER degradation. Note that the disciplined peak process is considered a kind of pulse shaping. In fact, there are several operations that can successfully mitigate the inter-symbol-interference (ISI) such that signal processing, coding, as well as decision feedback equalization methods [9]. Pulse shaping is considered one of the unexplored challenges in molecular communication systems until this moment. In fact, according to Brownian motion, the molecular channel response has a long tail which causes residual noise and Inter Symbol Interference (ISI) in the communication channels [1], [4].

In order to minimize these problems, similar to classical communication systems, it is suggested that a pulse shaping filter be used at the transmitter to minimize the ISI impact. Channel characteristics for the propagation medium must be studied in further detail before an appropriate pulse shaping filter can be selected. It can be shown that the function of the molecular concentration at the receiver in response to an impulse of molecular emission from the transmitter with Q molecules is of the form [8]:

$$U(t,d) = Q \frac{1}{(4\pi D t)^{\frac{3}{2}}} exp(-\frac{d^2}{4Dt}) \qquad (1)$$

where d denotes the distance between the receiver and the transmitter, and D is the diffusion constant. Note that the time at which the pulse has its maximum can be obtained from $\frac{\partial U}{\partial t} = 0$. Therefore, $t_{peak} = \frac{d^2}{6D}$.

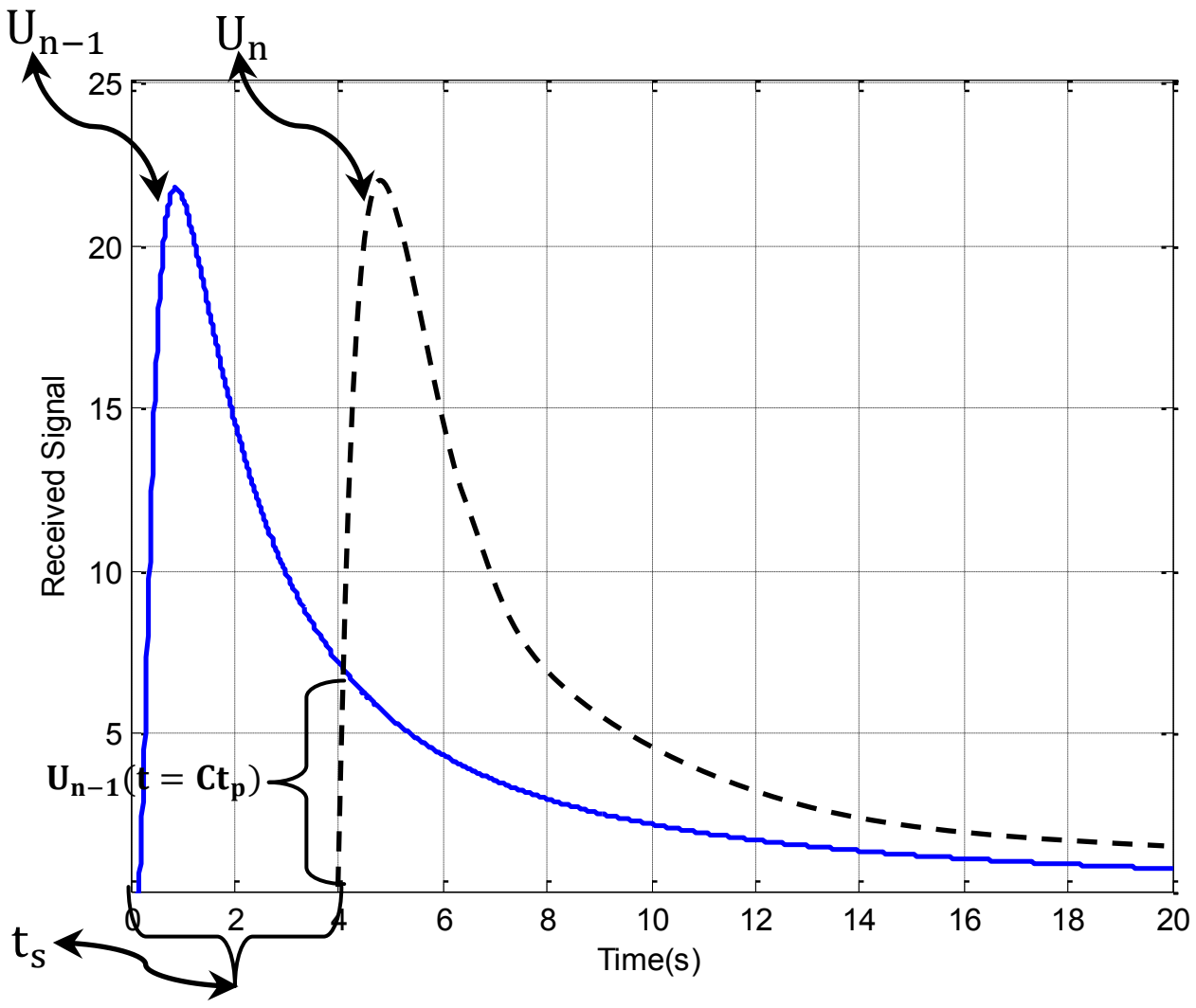

**FIG 2, HELPFUL ILLUSTRATION FOR DISCIPLINED PEAK PROCESS. THIS IS THE RECEIVED INSTANTIOUS IMPULSE, ACCORDING TO BROWNIAN MOTION AND NOISELESS CHANNEL**

The following equations will evidence how to control or reduce the transmitted signal Q based on the previous transmitted signals so that the peaks of the actual received signals will approximately remain constant at the receiver. The peak of the expected current received signal after DP process can be written as follows,

$$([U(t_p,d)]_n)_{after\ DP} = ([U(t_p,d)]_n)_{before\ DP} - [U(t_s + t_{p,},d)]_{n-1} - \dots\dots\dots - [U(mt_s + t_p,d)]_{n-m} \qquad (2)$$

Where $t_p$ is the peak time and $t_s = k\, t_p$ is the symbol duration; $k$ is a constant, set by the designers, and $k > 1$.

The goal of the pervious equation is to eliminate the ISI effect of the pervious received symbols on the peak of the current received symbol. By substituting equation 1 in equation 2, and then divide the outcome by $\frac{1}{(4\pi D t_p)^{\frac{3}{2}}} exp(-\frac{d^2}{4Dt_p})$; the current transmitted signal $Q_n$ could be written as a function of the previous transmissions as follows,

$$(Q_n)_{after\ DP} = (Q_n)_{before\ DP} - Q_{n-1}[(\frac{1}{k+1})^{\frac{3}{2}} exp\left(-\frac{d^2}{4Dt}(\frac{1}{k+1} - 1)\right)] - \tau_{n-2} - \cdots - \tau_{n-m} \qquad (3)$$

$$\tau_{n-m} = Q_{n-m}\,[(\frac{1}{mk+1})^{\frac{3}{2}} exp\left(-\frac{d^2}{4Dt}(\frac{1}{mk+1} - 1)\right)] \qquad (4)$$

As shown in equation 3, it is important to note that the transmitter Tx requires to know the position of the receiver Rx to achieve the disciplined peak process. To perfectly achieve the disciplined peak process, $m$ is supposed to be too large. At the reception node, simple Sample-Based Detector (SD) would achieve the decoding mission by sensing the molecular channel at $t_p$ , $(k+1)t_p$ , $(2C+1)t_p$ , … , $(mC+1)t_p$. It is also a must for the receiver to know the distance between $T_x$ and $R_x$ in order to accurately determine $t_p$ , and $C_{max}$ ;*"see equation 5"*; so that the receiver can correctly decode the information symbols.

It is important to note that, the diffusion coefficient of the medium will have no effect on the attenuation of the molecular pulses throughout space as shown in equation (5).

$$C_{max} = U\,(d,t)|_{t=t_p} = (\frac{3}{2\,\pi e})^{3/2}\frac{Q}{d^3} \qquad (5)$$

In BCSK technique, a number of molecules is emitted in an instantaneous fashion by the transmitter to signify logical 1; no molecule is emitted to signify 0, i.e., $(Q_n)_{after\ DP-logic\ "0"} = 0$. The decoding process of this scheme is so naïve for the BCSK scheme, i.e., the receiver senses the molecular channel at $t_p$ , $(k+1)t_p$ , $(2k+1)t_p$ , … , $(mk+1)t_p$ ; if the concentration of the molecular is equal or exceed $C_{max}$ then it's logical "1", otherwise its logic "0". The reception node is assumed to be perfectly synchronized with the transmission node.

It is important to note that if k is large enough then $m$ should not be large; however, if we would like to maximize the data rate i.e. small k, therefore, a lot of previous emissions must be taken into account.

From our observation and according to the following parameters (BCSK scheme, D= 0.43 $cm^2$/second, distance between $T_x$ and $R_x$ = 1.5 cm, and $Q_{before\ DP}$ = 1000 molecules), we conclude that the information of 20 previous emissions and 40 previous emissions are sufficient for k=4 and k=2 respectively, to produce zero bit error rate for 1000 random generating bits in a noiseless channel.

It is important to note that the maximum computational time of $(Q_n)_{after\ DP}$ cannot exceed the symbol duration, hence, if k is very small then the transmitter will have a small period of time to compute the optimal $(Q_n)_{after\ DP}$. In fact, the computational complexity of the proposed transmitters might surpass what could be implemented on the first generation of nano-machines.

In a further study, we will present some ideas to reduce the complexity of the discipline peak process, and we will evaluate the performance of the various CSK schemes in noisy channels. It is interesting to note that, according to the DP process, the decoding process of the M-ary Concentration Shift Keying (M-CSK) will be much easier than ever before since the peaks remain disciplined.

To summarize, the advantages of discipline peak process are: it consumes less no. of molecules per message; it maximizes the data rate, it suppresses the ISI effect; and its receiver requires very simple detector. However, the main disadvantage is: it requires complex transmitters.

## III. Orthogonal Molecular Division Multiplexing

Orthogonal Molecular Division Multiplexing (OMDM) is analogous to Orthogonal Frequency Division Multiplexing (OFDM) in conventional communication systems since it simultaneously utilizes several orthogonal sub-channels. For further explanation of OMDM, we will present a simple example of OMDM which is called Binary-Orthogonal Molecular Division Multiplexing (B-OMDM) and surpasses the performance of BMoSK in terms of channel capacity. B-OMDM might be described as follows:

### A. The transmitter has the following specifications:

(1) Has the ability to simultaneously emit two different types of molecules into two different orthogonal sub-channels.

(2) Has the ability to apply the disciplined peak process to each sub-channel.

(3) Modulate the information bits by using BCSK technique in each molecular sub-channel.

B. The Channel response and its specifications:

(1) The emitted molecules are diffused in a three dimensional space. The molecules are assumed to diffuse freely, and the dynamics is described by the Brownian motion.

(2) The molecular channel is assumed to be noiseless.

C. The receiver has the following specifications:

(1) The receiver has several receptors to detect the several kinds of molecules; each receptor can only detect one certain kind of molecules i.e. the receptor is equipped with a specific detector for each molecule type, the signal components (the different molecule types) are orthogonal interfere with each other.

(2) The symbol durations are equal i.e. $t_{s1}$=$k_1 t_{p1}$=$t_{s2}$=$k_2 t_{p2}$. Since each type of molecules has different diffusion factor, hence, it's recommended to choose the molecules which has close diffusion factor in the blood plasma, for instance, Glucose and Galactose in blood plasma. Note that $k_2 = \frac{k_1 D_2}{D_1}$ and $k_1$ is a constant set by the designers.

(3) The receptors of type '1' have to sense the molecular sub-channel at $t_{p1}$ , $(k_1+1)t_{p1}$ , $(2k_1+1)t_{p1}$ , … , $(mk_1+1)t_{p1}$ and the receptors of type '2' has to sense the molecular sub-channel at $t_{p2}$ , $(k_2+1)t_{p2}$ , $(2k_2+1)t_{p2}$ , … , $(mk_2+1)t_{p2}$ . If the concentration of the molecular is equal or exceed $C_{max}$ then it's logical "1", otherwise, it's logical "0".

(4) The reception node is assumed to be perfectly synchronized with the transmission node.

To make a fine understanding and for auxiliary illustration, we consider the following example: Assume that "1001" is required to be delivered, therefore, the transmitter will encode the message into types 1 and 2 molecules as follows; simultaneously emits certain no. of molecules 1, and emits nothing; afterwards, wait a certain time called symbol duration and then simultaneously emits nothing; and emits certain no. of molecules 2.

At the reception, The receptors of type '1' have to sense the molecular sub-channel at $t_{p1}$ , $(k_1+1)t_{p1}$ and the receptors of type '2' has to sense the molecular sub-channel at $t_{p2}$ , $(k_2+1)t_{p2}$. If the concentration of the molecular is equal or exceed $C_{max}$ then it's logical "1", otherwise, it's logical "0". Therefore, the first bit is logical one as well as the fourth bits, and the second bit is logical zero as well as the third bit. Note that B-OMDM doubles the channel capacity of BMoSK and consumes the same no. of molecules in case of Pr(bit '0')= Pr(bit'1').

It is interesting to note that B-OMDM serves the purpose of the paper, i.e., boosts the throughput and minimizes the emitted no. of molecules with respect to B-MoSK. Unfortunately, we cannot always achieve the purpose of the paper, mainly because if we would like to maximize the aggregated throughput of a network, then the transmitters will require emitting a large no. of molecules. The following discussion will illustrate the previous fact from network engineering prospective.

It is so essential when designing messenger molecules for biomedical applications to choose them to be non-toxic and non-harmful to human body. The messenger molecule suggested in [7], however, is toxic and flammable, which means it may be improper as a messenger molecule. Hexoses group is not harmful to human-body at all, and it has a total of 32 different isomers. In [10], hexoses group is utilized to create 32-IMoSK modulation scheme, which has the ability to convey 5 bits per symbol.

There is limited no. of non-harmful messenger molecules to human body, so that considering all the possible isomers for a pair of nano-machines will prevent the networks' designers from building a reasonable network. As mentioned in section I, there are only two possible accessing techniques for DMC networks, MDMA and TDMA.

MDMA technique uses different molecule types in order to perform several simultaneous transmissions, sharing the same medium but without interfering with each other. Assume a system that utilizes hexoses group (32 isomers) and the nano-machines use (B-OMDM) modulation scheme to covey the information; hence, we have 16 non-interfering channels in our network and each channel has a capacity of two bits.

However, if we consider 32-IMoSK modulation scheme, then we just have only one non-interfering channel in our network with a channel capacity of five bits. By deploying TDMA for the latter case, several nano-machines can share this channel where every nano-machine is allocated with a certain time slot.

From the previous discussion, it is obvious that the latter network scheme consumes much fewer molecules; however, the former network scheme provides much higher aggregated throughput.

## IV. Multi-hopping Gain

It is interesting to note that multi-hopping issue is so essential in the DMC networks, not only in delivering the information from the source to the destination but it has a great influence in maximizing the end-to-end throughput, reducing the overall consumed molecules Q and mitigating the interference. Note that the tails of the molecular channel response for the small distances is significantly smaller than the tails of the molecular channel response of the large distances.

In fact, the maximization of the end-to–end throughput is considered one of the beneficiaries of the multi-hopping mechanisms. Note that the end-to-end throughput is depending on the symbol duration, the no. of symbols per emission, and the no. of the hops per route. Moreover, $(k)$, the diffusion factor $(D)$, and the distance between the Source and Destination $(d)$, affect the end-to-end throughput since the symbol duration, $t_s = c\ t_p = \frac{cd^2}{6\text{D}}$. In general the governing equation in case of one hop might be written as follows,

$$\text{Th} = \frac{\text{n}}{t_s} = \frac{\text{n}}{kt_p} = \frac{6\text{nD}}{kd^2} \quad (\text{bit/s}) \tag{6}$$

Where; (n) represents the bandwidth efficiency of the molecular channel. By dividing the route into several hops (N) instead of one, therefore, the governing equation might be written as follows,

$$\text{Th} = \frac{6\text{nD}}{k(\frac{d}{N})^2 * N} = \frac{6\text{nD}*\text{N}}{kd^2} \quad (\text{bit/s}) \tag{7}$$

In fact, the route of (N) hops and distance (d) can convey the data (N) times faster than the route of one hop and distance (d). Note that each hop in the former case has a fixed length of $\frac{d}{N}$. Figure 6 visualizes the end-to-end throughput for the distance of 10μm (between the source and the destination) versus the no. of hops per route.

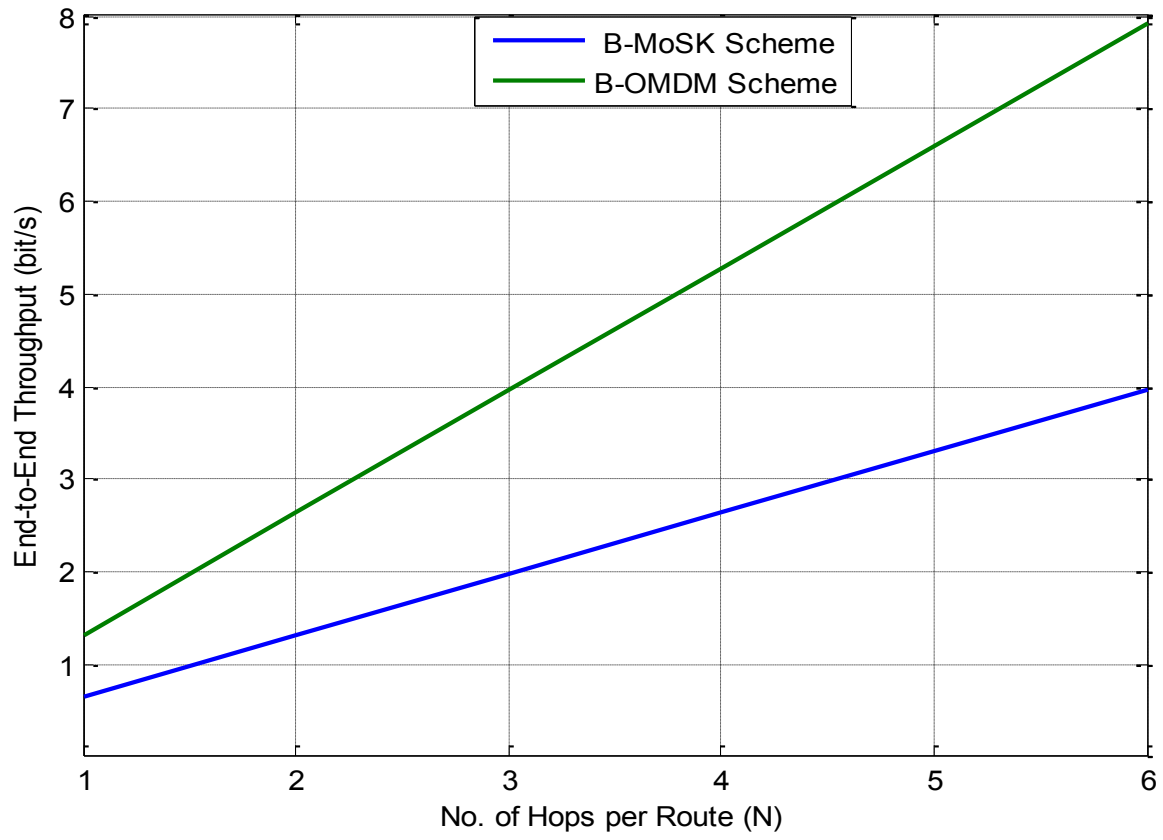

**FIG 3, The End-to-End throughput performance according to distance of 10 μm between the Source and the Destination. Average Diffusion Constant $(D)$ of 2.2X10$^{-7}$ cm$^2$/second and $(k)$ of 2 are utilized.**

Multi-hopping mechanisms can significantly reduce the overall consumed molecules Q in DMC networks. Obviously, this advantage is considered the most powerful beneficiary of the multi-hopping mechanisms. The following equations will evidence how the multi-hopping mechanisms can ultimately minimize the overall consumed molecules Q. In our study, we assume that $(C_{max})_{one\ hop}$ and $(C_{max})_{Multi}$ are equal. Substitute in equation 6, to obtain the following,

$$(\frac{3}{2\ \pi e})^{3/2} \frac{Q_{one\ hop}}{d^3} = (\frac{3}{2\ \pi e})^{3/2} \frac{N^3 Q_{Multi}}{d^3} \tag{8}$$

Note that $(C_{max})_{one\ hop}$ represents the peak concentration of the received molecular in case of one hop and distance (d), $(C_{max})_{Multi}$ represents the peak concentration of the received molecular in case of multi hops and distance $(\frac{d}{N})$, and (N) represents the no. of hops per route. Then, $Q_{one\ hop}$ can be written as a function of $Q_{Multi}$ as follows,

$$Q_{one\ hop} = N^3 Q_{Multi} \tag{9}$$

Interestingly, it's obvious that the no. of required emitted molecules to deliver a message through a route of (N) hops and distance (d) is $N^2$ times less than the overall required emitted molecules to deliver a message through a route of one hops and distance (d).Therefore, it is recommended to divide the route as long as possible in order to minimize the consumed Q molecules.

# V. Conclusion

In order to boost the throughput (the information data rate) with minimal amount of of transmitted molecules; disciplined peak process and Binary- Orthogonal Molecular Division Multiplexing (B- OMDM) are proposed; nevertheless, multi- hopping gain had been discussed. Despite of the computational complexity of the proposed techniques, we hold out hope that novel conscious computational methods will solve the problem. Profound investigations on how to connect the molecular communications networks tightly with electromagnetic networks will be offered in a further research.

# Transactional References